\documentclass[fleqn,10pt]{wlscirep}
\title{Nucleation and strain-stabilization during organic semiconductor thin film deposition}

\author[1]{Yang Li}
\author[1]{Jing Wan}
\author[2]{Detlef-M. Smilgies}
\author[3]{Nicole Bouffard}
\author[4]{Richard Sun}
\author[1,*]{Randall L. Headrick}
\affil[1]{Department of Physics and Materials Science Program, University of Vermont, Burlington VT 005, USA}
\affil[2]{Cornell High Energy Synchrotron Source, Cornell University, Ithaca NY 14853, USA}
\affil[3]{Microscopy Imaging Center, College of Medicine, University of Vermont, Burlington VT 05405, USA}
\affil[4]{Angstrom Sun Technologies Inc. 31 Nagog Park, Acton MA 01720, USA}

\affil[*]{rheadrick@uvm.edu}

\begin{abstract}
The nucleation mechanisms during solution deposition of organic semiconductor thin films determine the grain morphology and may influence the crystalline packing in some cases. Here, in-situ optical spectromicroscopy in reflection mode is used to study the growth mechanisms and thermal stability of 6,13-bis(trisopropylsilylethynyl)-pentacene  thin films. The results show that the films form in a supersaturated state before transforming to a solid film.  Molecular aggregates corresponding to subcritical nuclei in the crystallization process are inferred from optical spectroscopy measurements of the supersaturated region. Strain-free solid films exhibit a temperature-dependent  blue shift of optical absorption peaks due to a continuous thermally driven  change of the crystalline packing.   As crystalline films are cooled to ambient temperature they become strained although cracking of thicker films is observed, which allows the strain to partially relax. Below a critical thickness, cracking is not observed and  grazing incidence X-ray diffraction measurements confirm that the thinnest films are constrained to  the lattice constants corresponding to the temperature at which they were deposited.  Optical spectroscopy results show that the transition temperature between Form I  (room temperature phase) and Form II (high temperature phase) depends on the film thickness, and that Form I  can also be strain-stabilized up to  135$^\circ$C. 
\end{abstract}
\begin{document}

\flushbottom
\maketitle

\thispagestyle{empty}

\section*{Introduction}

Methods for solution deposition of organic semiconductor thin films have proven to be far superior to vapor deposition methods in many respects, such as the ability to create oriented films with extremely large crystalline grain size and exceptional carrier mobility.\cite{wo2008APL,minemawari2011inkjet,Ishviene2013JAP,diao2013NM,Niazi2015NatComm} However, the specific crystallization processes remain largely unknown, and no clear consensus has emerged on the mechanisms behind the stabilization of metastable polymorphs.\cite{Ishviene2013JAP,Diao2014JACS,Giri2014NatComm}  Classical nucleation theory posits that critical nuclei grow via attachment of monomers from supersaturated solution, while subcritical nuclei tend to thermally dissociate.\cite{Porter_and_Easterling}   However, there is abundant evidence that a two-step process, where crystals grow via attachment of pre-nucleation clusters,  takes place in  a diverse range of materials systems.\cite{Deniz2009ACR,gebauer2014pre,Ito:2016aa}  In addition, delayed crystallization from a supersaturated solution,\cite{wang2010development,shin2013effect,engmann2015real,chou2014late}  and evidence for aggregation that is not correlated to crystal diffraction\cite{duong2012role,Richter2015AEM,sanyal2011situ} have been observed in the context of drying of polymer thin films and polymer blends. Deposition on heated substrates poses an additional challenge since the stable state at the deposition temperature can become metastable once the temperature is changed, typically to ambient conditions.  Here, we seek to understand both of these critical steps in the solution deposition process $-$ nucleation and metastable state stabilization $-$  by performing a combination of in-situ and post-deposition characterization experiments.

It is widely appreciated that molecular packing strongly affects the electronic couplings between neighboring molecules in solid materials, resulting in substantial peak shifts in absorption spectra relative to isolated molecules.\cite{Kasha1965PAC,Spano2010ACR} However, there have been relatively few reports of in-situ optical spectroscopy to monitor dynamic processes during solidification of organic semiconductor thin films.\cite{peet2008transition,wang2010development,duong2014mechanism,Abdelsamie2014JMCC,Richter2015AEM}  Since the performance of electronic devices is heavily dependent on thin film morphology and structure, understanding and control of molecular self-assembly has significant practical implications. Moreover, understanding the effect of strain on the properties of small-molecule based organic semiconductors has recently emerged as an area of fundamental and practical importance.\cite{Wu:2016aa,Kubo:2016aa}

In this paper, we demonstrate the application of real-time optical spectroscopy to study the crystallization process and polymorphic transformation of  6,13-bis(triisopropylsilylethynyl) pentacene (TIPS-pentacene) thin films. TIPS-pentacene  is one of the most heavily studied prototypical organic semiconductors because of its solution processability and excellent device performance.  It has previously been reported that that the  $\pi-\pi^*$ absorption features of static TIPS-pentacene thin films are broadened and red shifted  compared to solution spectra by as much as 40 nm.\cite{James2013ACSNano,Oksana2005JAP}   We utilize in-situ optical spectroscopy to show that TIPS-pentacene crystallizes in a two-step process, where in the first step the film exists in a supersaturated form with no long range order, as reported previously based on in-situ synchrotron X-ray scattering and Polarized Optical Video Microscopy (POVM) results.\cite{Ishviene2013JAP}  We report evidence that TIPS-pentacene molecules aggregate into small clusters in the supersaturated form. Thus, molecular ordering begins in the first  stage of the film self-assembly process as a precursor to the development of long-range ordering.

\begin{figure}
\begin{center}
\includegraphics[width=5.0 in]{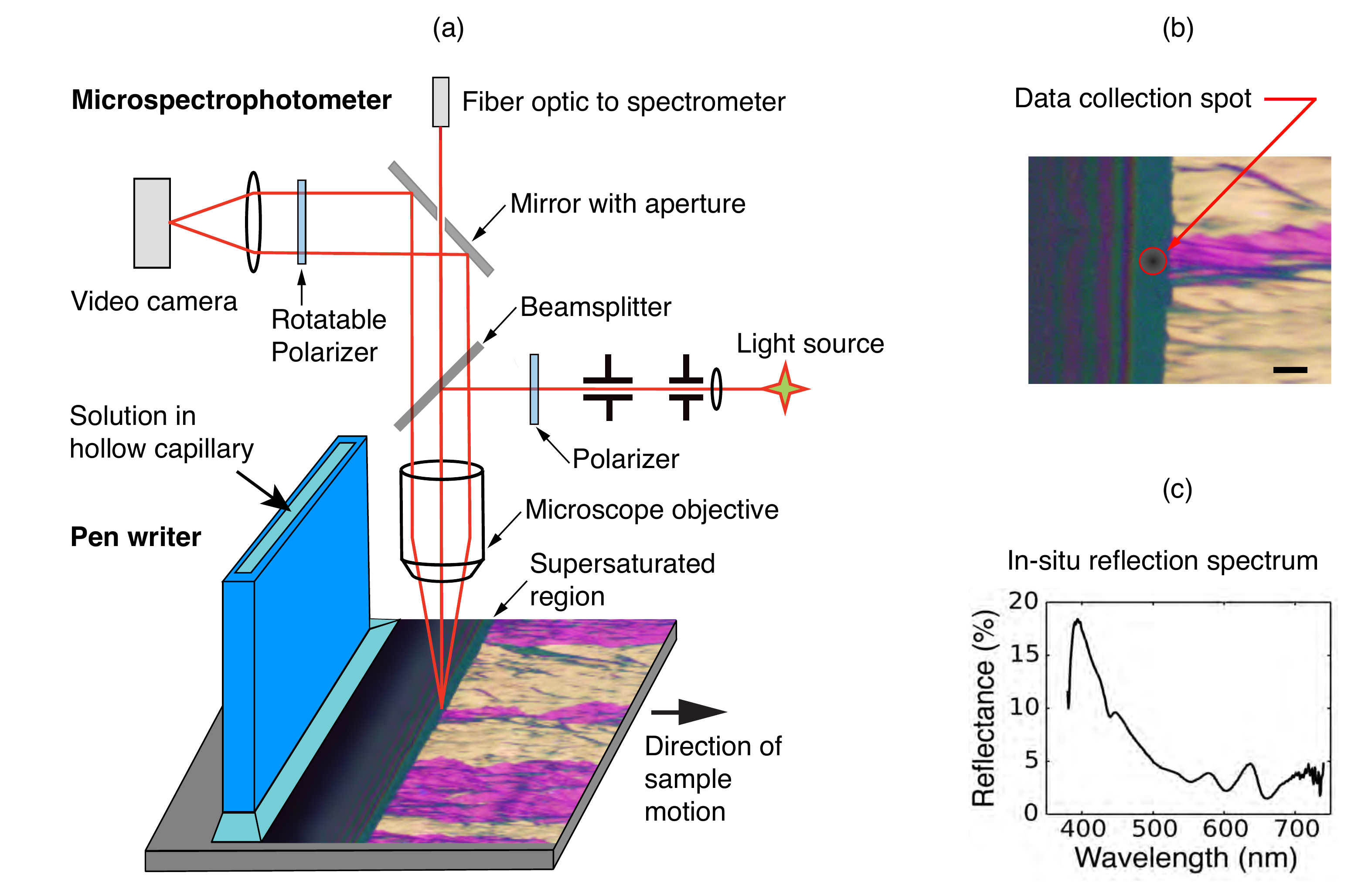}
\caption{\label{Expt_Layout} (a) Schematic of the real-time optical reflectance and microscope system integrated with the hollow capillary pen writer. A camera is used to monitor the growth process and a spectrometer is used to simultaneously collect reflection spectra. A mirror with an aperture is inserted into the light path to select a small area of interest, as indicated by the dark spot in (b).  The sample shown in the optical image is made from an  8.7 mg/ml TIPS-pentacene solution. The writing speed is 0.4 mm/s and the deposition temperature is 25$^\circ$C. This film is made on a Si/SiO$_2$ substrate and the rotatable polarizer placed at 45$^\circ$ in order to provide improved contrast on the grain structure. (c)  A typical real time reflection spectrum. For real-time reflectance, silicon is used as the substrate and the experiment is done without any polarizers. }
\end{center}
\end{figure}

Cour $et al.$\cite{Ishviene2013JAP}  have shown that strained small molecule organic semiconductor thin films can be obtained by solution deposition at elevated temperatures (60$^\circ$C and 90$^\circ$C for TIPS-pentacene).  The strain is introduced after the deposition when the sample temperature is changed, as a result of the very large thermal expansion coefficient of the solid film relative to typical substrate materials such as silicon and glass.\cite{Ishviene2013JAP} Highly strained films are obtained after cooling unless the strain is relieved by cracking or buckling.  In addition, TIPS-pentacene undergoes a bulk phase transition at 124$^\circ$C.\cite{Chen2006JPCB} It has also been reported that it is possible to stabilize thin films in the high temperature phase, known as Form II, to ambient conditions.\cite{diao2013NM}  Here, we report that in-situ optical spectroscopy is highly sensitive to changes in molecular packing and to thermal expansion effects.  We utilize this effect during thermal cycling in combination with X-ray diffraction measurements to verify that Form II TIPS-pentacene films deposited at 135$^\circ$C can be stabilized to room temperature if the film thickness is below the critical thickness for cracking.   We have also performed similar experiments on room temperature phase (Form I) TIPS-pentacene thin films written at 25$^\circ$C. No phase transition occurs even when the sample is heated to 135$^\circ$C, which provides  confirmation of the strain-stabilization model since no reduction of the transition temperature is observed.

\begin{figure}
\begin{center}
\includegraphics[width=5.25 in]{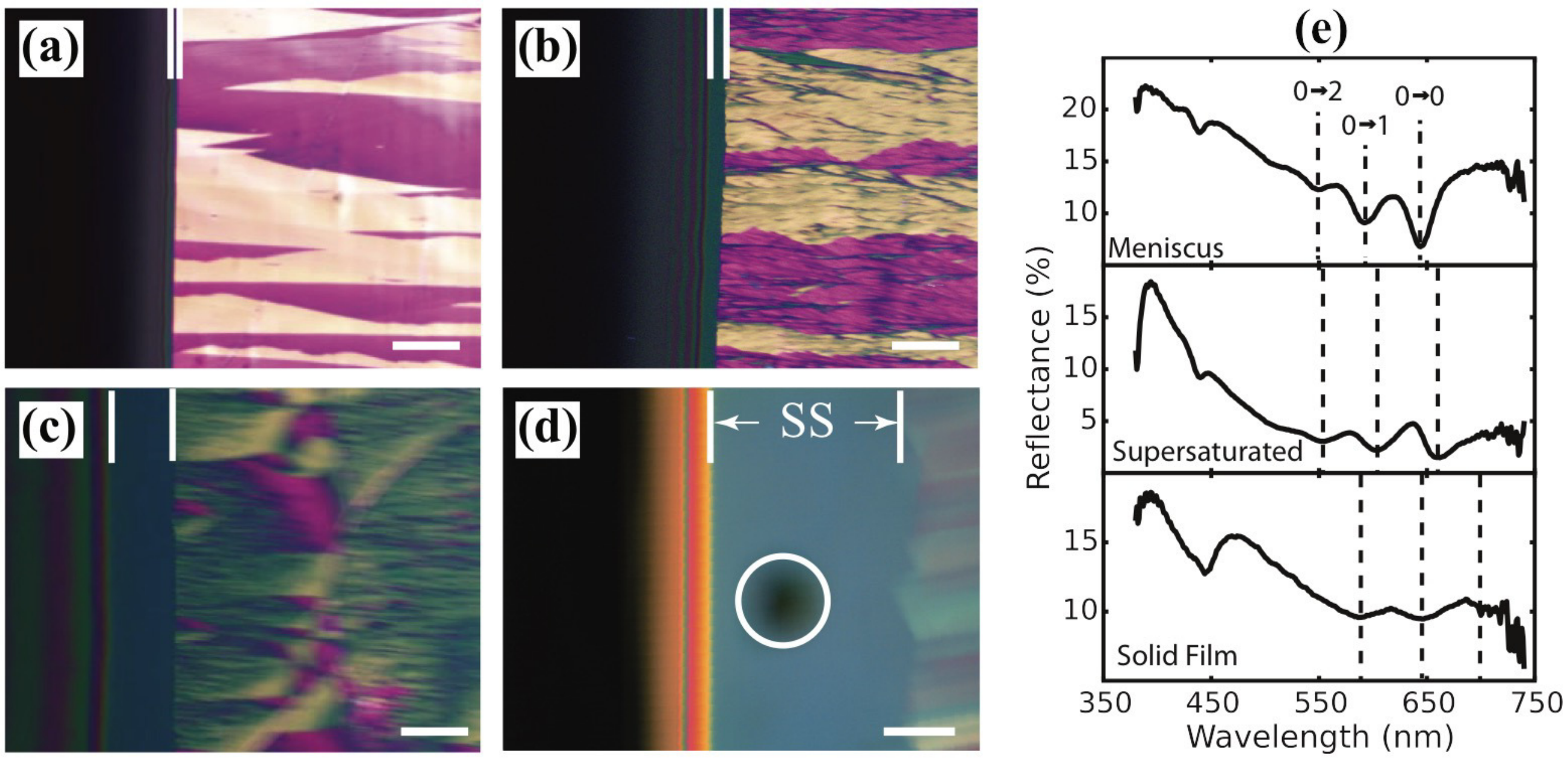}
\caption{\label{optical} Optical  microscopy images at different writing speeds: (a) 0.2 mm/s, (b) 0.4 mm/s, (c) 1 mm/s, (d) 0.4 mm/s at higher magnification.   The solution concentration was 8.7 mg/ml. From left to right, three regions can be seen during thin film deposition: meniscus, supersaturated region marked by vertical white lines, and solid film. In (a)--(c)  films were prepared at 25$^\circ$C on silicon wafers with 300 nm silicon dioxide layer, and the images was taken using a 45$^\circ$ polarizer.  In (d) a microspectrophotometer is used to collect real-time  reflection spectra from the darkened region indicated by a white circle. The conditions are similar to (b) except that a Si wafer was used as substrate and no polarizers were used.    The scale bars are 200 $\mu$m for (a)--(c), and 50 $\mu$m for (d). (e) Real-time reflectance spectra for three regions during solution deposition at 0.4 mm/s and 25$^\circ$C on Si (with no oxide layer). The final film thickness for this sample was $\approx$30 nm.}
\end{center}
 \end{figure}

\section*{Results and Discussion}

\subsection*{In situ optical monitoring of the crystallization process}

Wo \emph{et al.}\cite{wo_thesis_2011,wo2012JAP} have shown that the film morphology prepared by the hollow pen writing method is highly dependent on the writing speed. At low writing speed ($<$0.4 mm/s for room temperature deposition), the crystallization of TIPS-pentacene is an evaporation-induced process occurring at the contact line. Oriented and nearly single crystalline films can be produced because the crystallization is seeded by the solid film emerging from the meniscus. When the writing speed is increased above a critical value ($\approx$2 mm/s), the dynamic meniscus becomes much larger as the solution is dragged out of the capillary by viscous forces, leaving a wet film.  Then, the crystallization no longer occurs at a well-defined contact line.  Instead, nucleation occurs randomly to form a spherulitic grain structure.\cite{wo_thesis_2011,wo2012JAP,Ishviene2013JAP}  We note that the grain morphologies obtained in the different speed regimes can also be observed for films deposited by closely related deposition methods such as zone casting, blade coating, and solution shearing, indicating that the processes involved are general and not specific to hollow capillary pen writing.\cite{diao2014morphology}

       Fig. \ref{optical} shows results in the intermediate speed regime where the dynamic meniscus is beginning to stretch into a continuous wet film.  Three regions of the film are observed during the deposition process: (i) the meniscus showing color fringes due to the rapidly varying  thickness of the solution is at the left side of each image, (ii)  a narrow featureless region is visible near the middle,  and (iii)  the solid film  with crystalline grain structure.  The middle region, which we will refer to below as the supersaturated region  is completely dark in 90$^\circ$ cross-polarized microscopy and does not exhibit any color fringes, indicating that it is a very thin isotropic layer. The concentration of TIPS-Pentacene in this layer is much larger than the solubility limit ($\approx$50 mg/ml in toluene at 25$^\circ$C) since the layer thickness is indistinguishable from the thickness of the final solid film. The width of this region increases from 22  $\mu$m at a writing speed of 0.2 mm/s to 59  $\mu$m at 0.4 mm/s, to 198  $\mu$m at 1 mm/s as shown in Fig. \ref{optical}(a-c).  We can place the entire reflectometer data collection spot in the supersaturated region using a 20$\times$ objective lens when the writing speed is 0.4 mm/s, as shown in Fig. \ref{optical}(d). Thus, we are able to obtain reflection spectra for the supersaturated region and compare them to the reflection spectra of both the meniscus and the solid thin film. Note  that we move the substrate rather than the pen for these measurements so that the meniscus and supersaturated regions are always stationary relative to the microscope objective, as shown in Fig. \ref{Expt_Layout}. It is necessary to perform the experiment in this mode because the time to collect a complete optical spectrum is rather long (1 s.). We also observe that the grain size becomes smaller and less oriented as we approach the critical writing speed, in agreement with previous results.\cite{wo2012JAP}

\begin{figure*}
\begin{center}
\includegraphics[width=4.80 in]{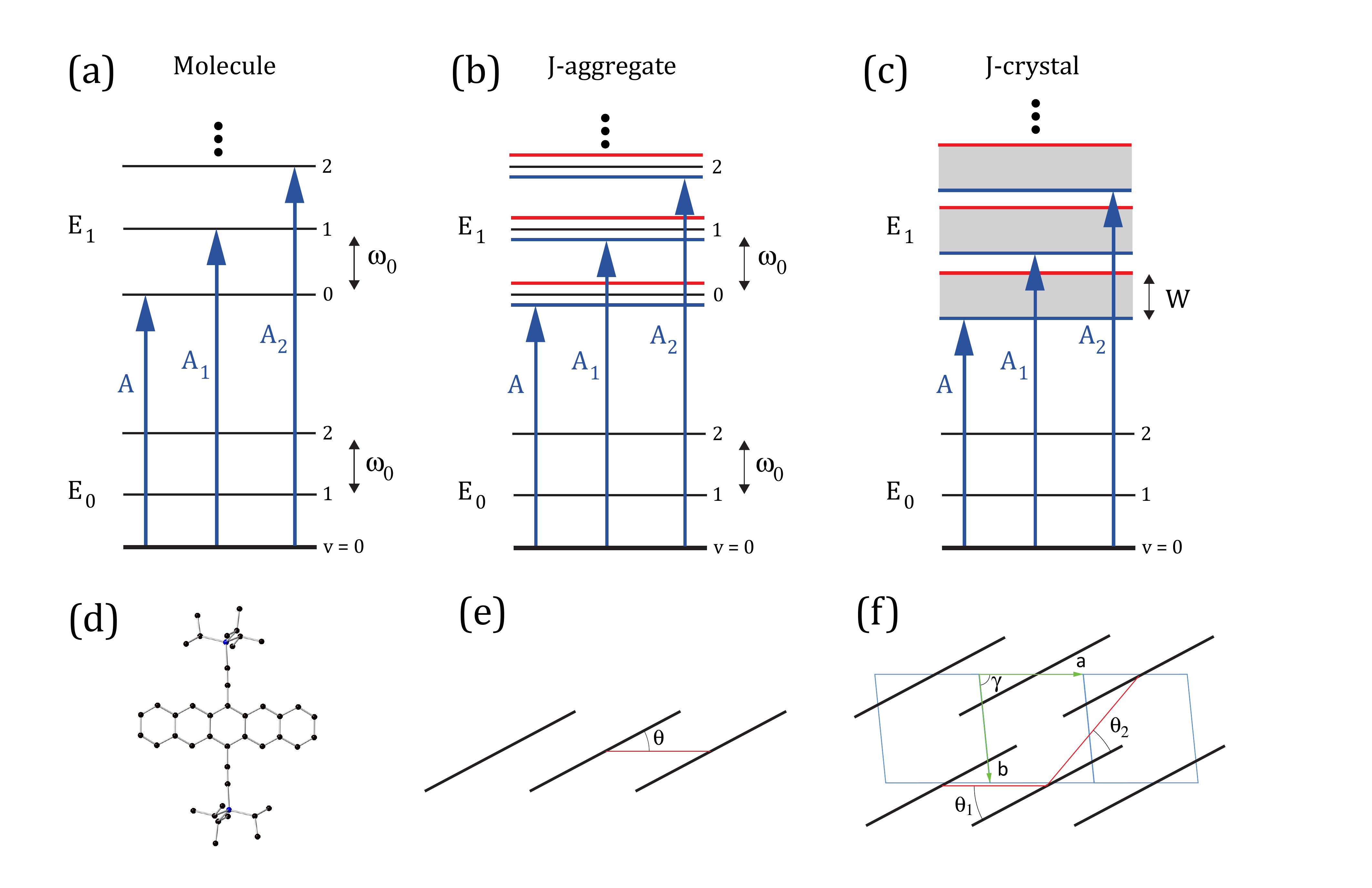}
\caption{\label{J-aggregate} Energy level diagrams for (a) a molecule, (b) a J-aggregate and (c) a J-crystal. The corresponding structural diagrams are: (d)  a TIPS-pentacene molecule, (e) the molecular arrangement of a TIPS-pentacene aggregate, and (f) the TIPS-pentacene crystalline unit cell. Excited states (E$_1$) split when molecules form aggregates (b) and become bands in the crystalline form (c).  When the TIPS-pentacene forms a crystal, each molecule interacts with its near neighbors; the angles $\theta$$_1$ and $\theta$$_2$ shown in (f) both affect the absorption peak position. Note that in (e) and (f) only the packing of the molecular cores are represented for simplicity; the side groups above and below the layer are not shown.}
\end{center}
\end{figure*}

 Typical reflection spectra for each region are shown in Fig. \ref{optical}(e).  Since we plot these spectra as reflection percentage, absorption features appear as downward peaks. Three peaks at 549 nm, 593 nm and 644 nm are observed for the meniscus spectrum, which is characteristic of the single molecule $\pi$-$\pi$* exciton.  The three peaks form a Frank-Condon series due to vibronic coupling in the molecule,  which  gives rise to a progression of absorption features labeled in Fig. \ref{optical}(e) as 0$\rightarrow$v (v = 0, 1, 2, ...) where v is a vibrational quantum number in the E$_1$ excited electronic state.\cite{yamagata2014JPCC}  Fringes from thin film interference effects are sometimes observed in these spectra, but are generally too weak to significantly shift the peak positions of the vibronic absorption features.  The E$_0$$\rightarrow$E$_1$ transition absorption peaks broaden and red shift as the solution drys and crystallizes. The red shifts for 0$\rightarrow$2, 0$\rightarrow$1 and 0$\rightarrow$0 absorption peaks in the supersaturated region are 5 nm, 11 nm and 15 nm respectively; additional details are shown in Supplementary Table S1.  There is further broadening and a large red shift when the supersaturated region transforms to solid. The red shifts for the  solid film are 33 nm, 51 nm and 56 nm for 0$\rightarrow$2, 0$\rightarrow$1 and 0$\rightarrow$0 absorption peaks respectively. A peak at $\approx$440 nm is observed in all three spectra, which has been assigned to an intramolecular excitation.\cite{James2013ACSNano}  Since these measurements are in  reflection mode, the peak positions may be shifted from the absorption positions by dispersion of the real part of the refractive index.  However, in Fig \ref{optical}(e) and Supplementary Table 1, the solid film reflection peak positions are within 5 nm of the corresponding absorption peak positions for reference films on glass measured in transmission (see Supplementary Fig. S1), indicating that dispersion effects are relatively small for the measurements on silicon substrates.

 Excitonic coupling  is very sensitive to the molecular packing and modifies the optical and photo-physical properties of the material. When two transition dipoles are side by side, their relative orientation causes absorption peaks to shift, and this effect can be classified into either J-type or H-type depending on whether the dipoles are head-to-tail  [$\theta < 54^\circ$ in Fig. \ref{J-aggregate}(e)] or  face-to-face, respectively. In the case of molecular aggregates (e.g. dimers or trimers), the excitonic interaction between the molecules results in splitting of the excited states.  For J-aggregates,  an optical transition to the lower excited state is allowed as shown in Fig. \ref{J-aggregate}(b). Thus, a red shift of absorption peaks is expected for J-aggregates.   TIPS-pentacene in crystalline form assume a 2D brickwork packing.\cite{Anthony2001JACS}  Each TIPS-pentacene molecule interacts with its  nearest neighbor molecules [Fig. \ref{J-aggregate}(f)]. Since the molecules are packed in a head-to-tail configuration with both $\theta_1$ and $\theta_2 < 54^\circ$, the crystal is itself expected to have the characteristics of J-type excitonic coupling.  However, the larger number of molecules in the crystal causes the excited states to form bands [Fig. \ref{J-aggregate}(c)].  
 
The aggregates in the supersaturated region are smaller than the critical nucleus size so that -- in both classical and two-step theories -- they continually  form and dissolve.\cite{vekilov2010} Since we observe minimal broadening of the absorption features, we deduce that the majority of the TIPS-pentacene molecules are actually incorporated into aggregates at this stage. We also find that in-situ X-ray scattering results are consistent with the solvent having almost completely evaporated as the the long-range ordering begins to develop, as shown in Supplementary Fig. S2. Therefore, it is likely that the crystallization process proceeds by incorporation of aggregates instead of from  individual molecules.  We infer that as the crystal is built up from these larger units, they may require some time to become fully integrated with the crystal.  This may at least partially account for the slow increase observed in X-ray diffraction peak intensities after the initial formation of a solid film.

 \subsection*{\label{sec:strain_free} Strain-free optical spectra vs. deposition temperature}

\begin{figure*}
\begin{center}
\includegraphics[width=4.25 in]{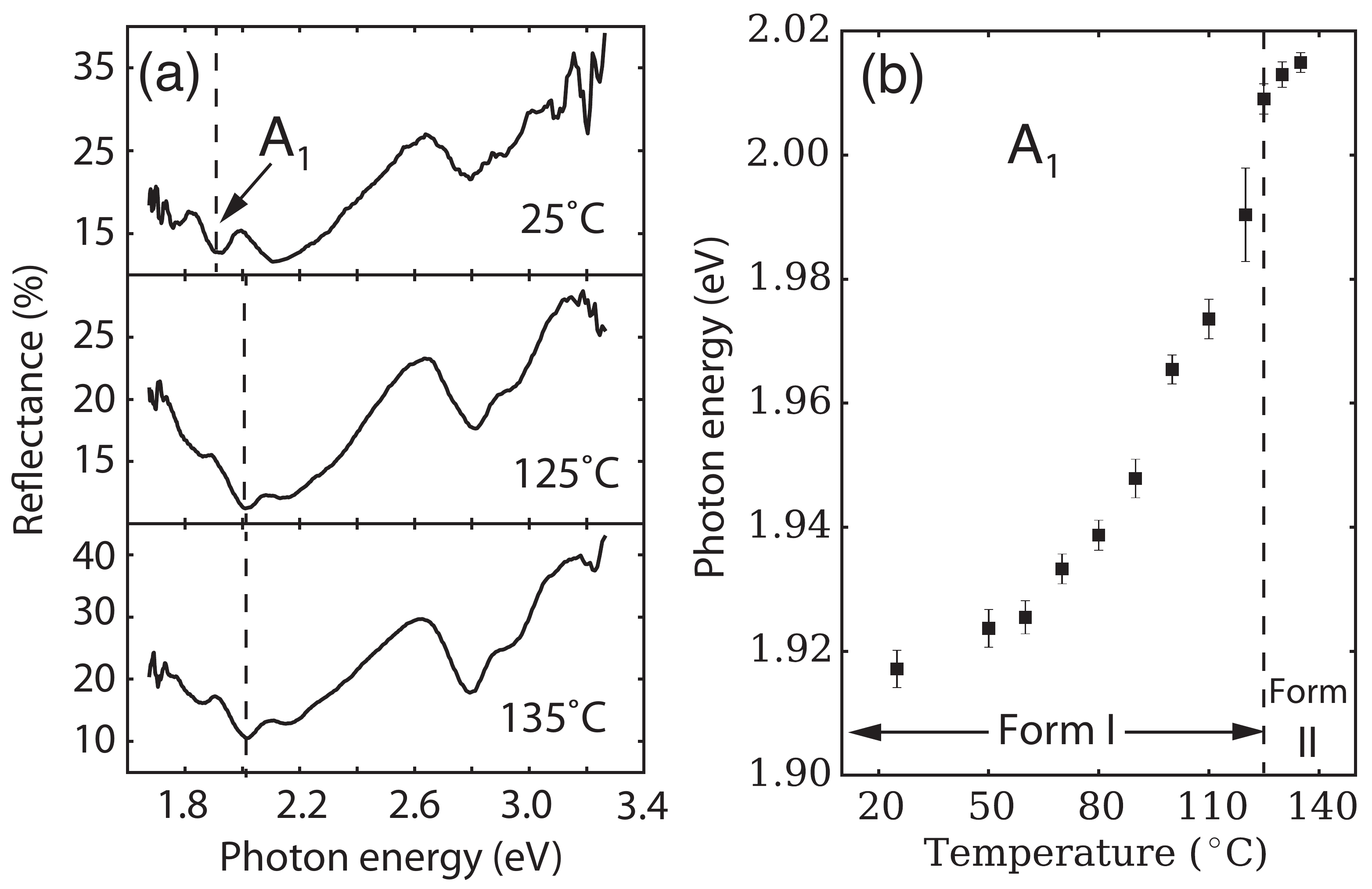}
\caption{\label{equilibrium_peak_position} In-situ reflection spectroscopy monitoring of TIPS-pentacene thin films prepared as a function of temperature from 25$^\circ$C to 135$^\circ$C. The reflection spectra were measured at the deposition temperature of each film. (a) reflection spectra collected at 25$^\circ$C, 125$^\circ$C and 135$^\circ$C. (b) the peak shift of the reflectance feature corresponding to the 0$\rightarrow$1 absorption peak (A$_1$) versus deposition temperature. The writing speed was kept at 0.05 mm/s. Glass slides were used as substrates and the concentration of solution was 1.5 mg/ml.  By convention, the films made above 125$^\circ$C are called Form II and films deposited below 125$^\circ$C are called Form I.}
\end{center}
\end{figure*}

It is well known that strain in TIPS-pentacene  thin films can lead to thermal cracking.\cite{chen_cracking_2006,Ishviene2013JAP}  As we have discussed above, the strain is introduced upon changing the temperature of solid films due to thermal expansion or by a phase transformation.  We anticipate that strain can cause noticeable peak shifts in absorption spectra. Therefore, we have carried out a study of the optical spectra of freshly made crystalline films recorded at their deposition temperatures in order determine the \emph{intrinsic unstrained} optical absorption peak position at each temperature.   We also utilize relatively thick films ($>$100 nm) in order to minimize the potential influence of surface or interface confinement effects on the molecular packing (film thickness versus temperature is shown in Supplementary Fig. S3). These results provide a reference for additional measurements on strained layers as a function of film thickness, which we report below.  Fig. \ref{equilibrium_peak_position}(a) shows examples of spectra covering the entire range of the low temperature phase (Form I) from room temperature up to and beyond the bulk transition to Form II at 124$^\circ$C. We plot the spectra as a function of photon energy instead of wavelength in order to more easily compare to the energy diagram shown in Fig. \ref{J-aggregate}(c). 

Since the reflection feature associated with the  0$\rightarrow$1 absorption peak (A$_1$)  is the strongest among the three lowest energy features, this peak height was chosen to study the structural evolution, as shown in Fig. \ref{equilibrium_peak_position}(b) (additional results for peaks A$_0$ and A$_2$ are shown in Supplementary Fig. S4).   The most pronounced features of this plot are the blue shift of the reflection features as the temperature is increased and the excellent sensitivity of the method to deposition temperature changes as small as 10$^\circ$C.    A blue shift  -- that is, a shift towards higher energy --  in the solid state absorption features generally corresponds to a narrowing of the bandwidth, denoted as $W$ in Fig. \ref{J-aggregate}(c).  The results are in good agreement with X-ray diffraction results for bulk powder samples if we interpret the temperature dependence as being related to thermal expansion effects.   For example, the  thermal expansion coefficient  is known to increase for TIPS-pentacene as the temperature is increased in the Form I,\cite{Chen2006JPCB} which is in agreement with the increasing slope of the data in Fig. \ref{equilibrium_peak_position}(b) as the temperature increases. It is interesting that there is a sudden change of slope at the bulk transition temperature, which may signal the phase change.  Chen $et al.$\cite{Chen2006JPCB} have speculated that the phase transition involves the rearrangement of the two bulky side groups of the TIPS-pentacene molecule.   Therefore, since the optical absorption features are mainly sensitive to the packing of  the molecular cores as we have discussed above, it is not surprising that there is no sudden jump at the transition temperature, which one would normally expect for a first order phase transition. The lack of first-order transition behavior in the packing of the core molecules is a key result, since it helps to explain why we can sweep the temperature of samples through the phase transition without re-nucleating the grain structure, which we do not observe (See the following two subsections).   
      
\subsection*{\label{sec:cracking}Cracking of films during cooling vs. film thickness}

\begin{figure}
\begin{center}
\includegraphics[width=5.5 in]{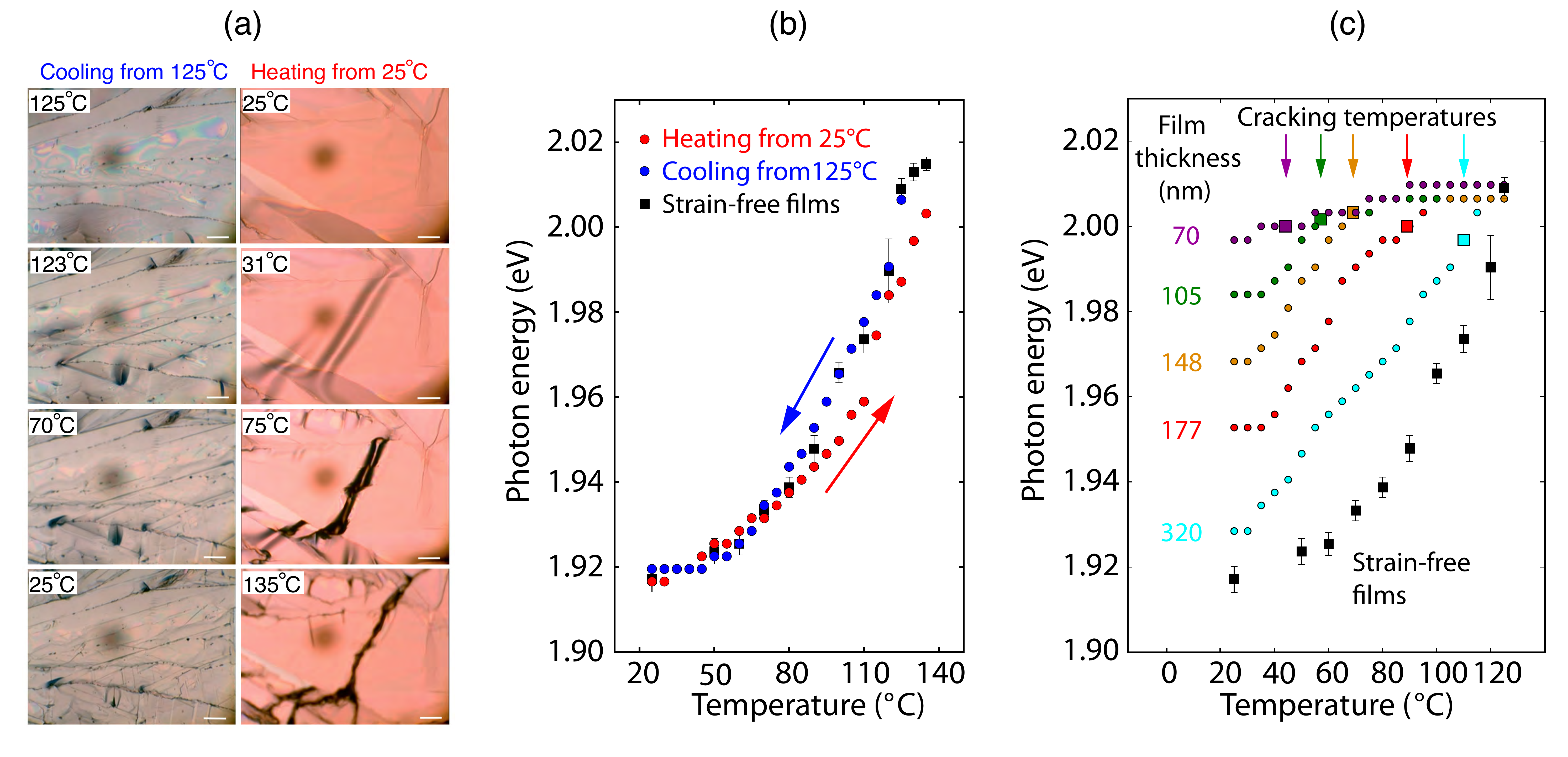}
\caption{\label{thick_film_cracking} (a) Comparison of heating and cooling for thicker films. The left sequence shows the deformation and cracking of a thick film made at 125$^\circ$C during cooling (740 nm, see Table \ref{tab:cracking_temperatures}), and the sequence on the right shows buckling and cracking of a thick film made at 25$^\circ$C during heating  (10 mg/ml; 0.02 mm/s; thickness was 230 nm; glass substrate).   The scale bars are 50 $\mu$m. (b) In-situ optical reflectance spectroscopy A$_1$ peak positions for each sample as a function of temperature. Blue circles: film made at 125$^\circ$C   and then cooled to 25$^\circ$C. Red circles: film made at 25$^\circ$C  and then heated up to 135$^\circ$. Black squares: absorption peak positions of unstrained films reproduced from Fig. \ref{equilibrium_peak_position}. (c) Thickness dependence of real-time optical reflection peak shifts in the cooling process. Samples with different thickness were made at 125$^\circ$C on glass slides and then cooled down to 25$^\circ$C.   The film thicknesses, writing speeds, and cracking temperatures are shown in Table \ref{tab:cracking_temperatures}. The black square symbols with error bars show the peak position for strain-free films prepared at different temperatures reproduced from Fig. \ref{equilibrium_peak_position}. The colored square markers and arrows indicate the temperatures that cracks start to appear.   }
\end{center}
\end{figure}

\begin{table}[ht]
\begin{center}
\caption{\label{tab:cracking_temperatures}Cracking temperatures versus film thickness for films written at 125$^\circ$C on glass slides. The concentration of the solution was  2.1 mg/ml in mesitylene. }
\begin{tabular*}{0.60\textwidth}{@{\extracolsep{\fill}}cccc}
 \hline \hline
 Writing speed (mm/s) & d (nm) & T$_{cr}$ ($^\circ$C) & $\Delta$T$_{cr}$ ($^\circ$C) \\
\hline
0.02 & 740 &  123 &    2 \\
0.05 & 320 &  110 &   15 \\
0.08 & 177 &  89   &   36 \\
0.1   & 148 &  69   &   56  \\
0.2   & 105 &  57   &   68  \\
0.3   &   70 &  44   &   81  \\
0.4   &   44 &  $- $ & $>$100$^a$  \\
 \hline \hline
\end{tabular*}
\begin{tabular*}{0.60\textwidth}{@{\extracolsep{\fill}}cccc}
$^a$ This film did not crack.  
\end{tabular*}
\end{center}
\end{table}

In previous work, strained lattices were obtained by depositing TIPS-pentacene thin films at elevated temperatures (60$^\circ$C and 90$^\circ$C).\cite{Ishviene2013JAP}  When thin films are constrained to  a substrate with a much smaller coefficient of thermal expansion, elastic energy builds up during cooling of the sample, and it relaxes as free edges are created through cracking.   Since the elastic energy is proportional to the film thickness, thinner films can accommodate a higher strain before failing.  In principle, films that are not entirely continuous can release strain without cracking. However, we find that in practice cracks generally appear with a final spacing of a few microns, so that pinholes and even stripe-like morphologies have a minimal effect on the cracking behavior.

We have prepared films with different thicknesses at 125$^\circ$C and then cooled them to 25$^\circ$C while acquiring in-situ optical spectra during the cooling process.  In-situ optical spectroscopy results for the cooling process are shown in Fig.  \ref{thick_film_cracking}, and    Table I summarizes the writing conditions, film thicknesses, and the cracking  temperatures.   We can see in Fig.  \ref{thick_film_cracking}(a) that the 740 nm thick film cracks almost immediately during cooling. Our strain model predicts that when the thickness is large, the strain energy quickly exceeds the cracking threshold, thus causing the reflectance peaks to shift towards their strain-relaxed positions. Consequently,  the optical data for this sample nearly tracks the behavior of the unstrained films, as we observed in Fig.  \ref{thick_film_cracking}(b). 

We can distinguish between a nucleation driven phase transformation and strain relaxation  by the sequence of cracking. In the first case, the cracks propagate outward from a nucleation point, as we have recently observed during crystallization of  C$_8$-BTBT (2,7-Dioctyl[1]benzothieno[3,2-b][1]benzothiophene) from solution.\cite{wan2016transient} In the second case, cracks appear with widely distributed positions and then increase in density, i.e. there is no obvious center of nucleation. As we can see in Fig. \ref{thick_film_cracking}, the cracking appears progressively and the grain structure is unchanged, suggesting that the cracking is due to strain. Fig. \ref{thick_film_cracking} also shows results for a thick film prepared at 25$^\circ$C and then subsequently heated to 125$^\circ$C.  During heating  the films buckle  due to compression rather than cracking from tension.  Buckling also leads to cracking and also relaxes strain so that we do not observe a large hysteresis between the heating and cooling curves in Fig. \ref{thick_film_cracking}(b).

The absorption peak positions versus temperature during cooling  for several additional samples listed in Table \ref{tab:cracking_temperatures} are shown in Fig. \ref{thick_film_cracking}(c).   The peak positions are all nearly the same at the starting temperature in spite the large range of thicknesses because they  are  initially unstrained.  A strong correlation between cracking of the film and the observed shifting of the absorption peaks can be seen.  This behavior is  consistent with the strain model, which predicts that thinner films can accommodate a higher strain, and thus the cracking  temperature is depressed further and further as the film thickness is reduced.    Moreover, the optical peak positions  relax towards the unstrained positions as the  temperature is lowered once the temperature is decreased below the cracking temperature.  However,  they shift gradually rather than abruptly and the peak width does not broaden appreciably. This suggest that  the film distributes strain relaxation over a region roughly comparable to the crack spacing.

\begin{figure}
\begin{center}
\includegraphics[width=4.5 in]{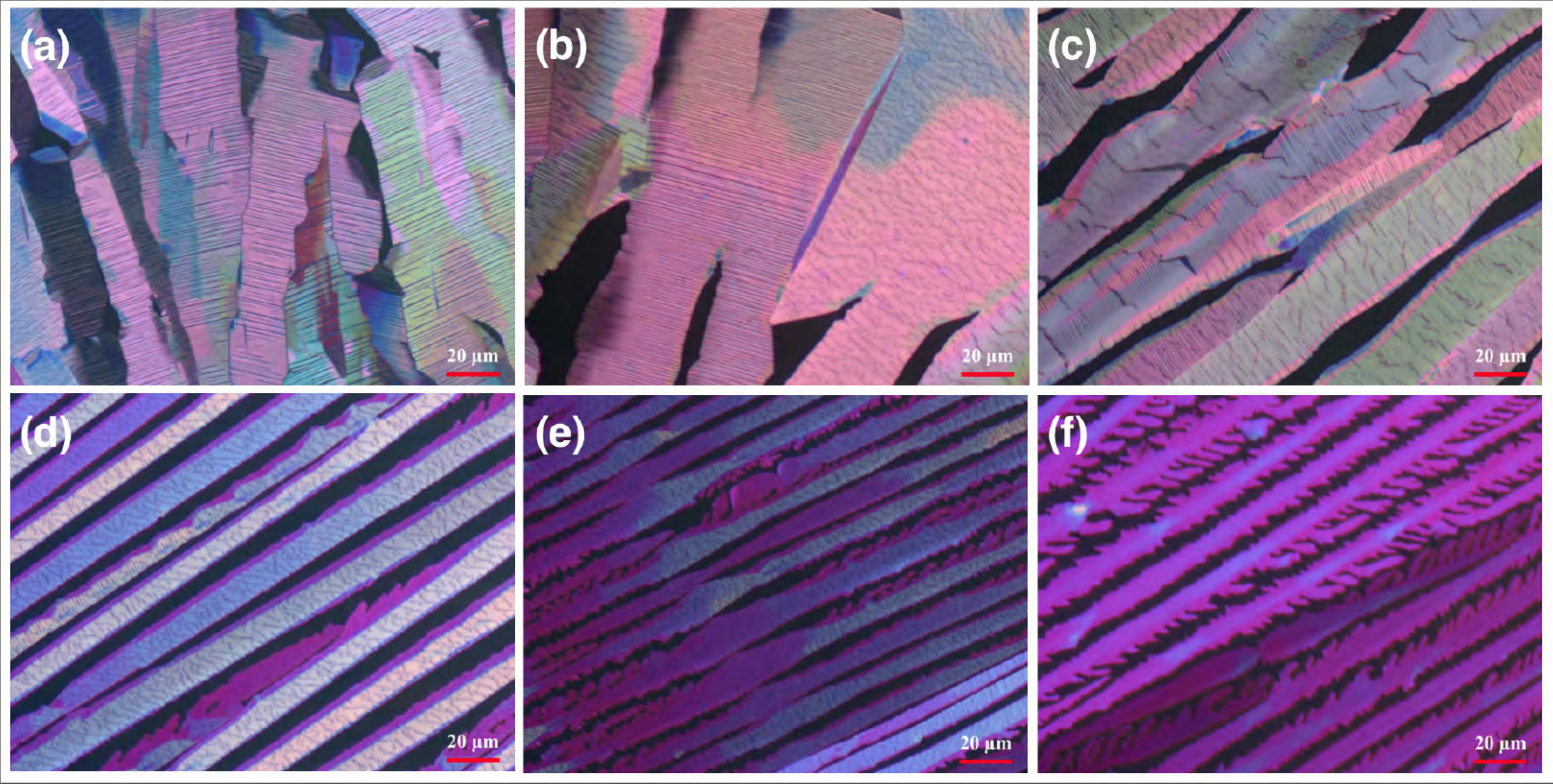}
\caption{\label{cracking_images} Polarized optical microscopy images of films prepared at 125$^\circ$C and then cooled to room temperature.  The  writing speeds in (a)—(f) are 0.05, 0.08, 0.1, 0.2, 0.3 and 0.4 mm/s respectively and the corresponding film thicknesses are 266, 164, 134, 95, 64 and 38 nm.  The writing direction is along the long axis of the grains in each case. The film deposition conditions are identical to those given in Table \ref{tab:cracking_temperatures}, except that they were deposited  onto Si/SiO$_2$ substrates instead of glass slides in order to improve the optical contrast to grain structure and defects. }
\end{center}
\end{figure}

      The final film morphologies of a series of samples deposted at 125$^\circ$C and then cooled to room temperature are shown in Fig. \ref{cracking_images}. We note that the results closely reproduce the film thicknesses and grain morphologies of the films deposited on glass described above, but the thermally oxidized Si substrates used for this set of samples produce higher quality optical images owing to the reflection from the substrate itself, as opposed to glass which reflects very little light. The images reveal that the density and size of the cracks changes as the film thickness is reduced.   The crack directions are mostly perpendicular to the writing direction (the stripe direction, close to the a-axis of TIPS-pentacene), consistent with the fact that the films tend to contract along the  a-axis by a large amount during the cooling due to the large thermal expansion coefficient along that direction. For the thicker films (266 and 164 nm), numerous straight features are observed that do not appear to be cracks. Rather, we speculate that they are deformation-induced twins since they exhibit polarization contrast that is rotated with respect to the orientation of the surrounding grains.  As the film thickness is reduced (134 and 95 nm) the straight features are gradually replaced by  more jagged,  open cracks.  This pattern persists at 64 nm, except in a few thinner spots. Finally, when the film thickness is 38 nm, cracks are entirely absent.  
      
      These results strongly support the strain model since the data shows that the films are strained, and that deformation and cracking are a result of strain exceeding a critical limit.  One could also invoke possible interface energetic effects or confinement effects to qualitatively explain some of the phenomena. However, strain is a volume effect, and as such  it is expected to strongly dominate over interface effects for layers more than 1 to 2 unit cells in thickness.

\subsection*{\label{sec:polymorph}Metastable polymorph fabrication and stabilization}

\begin{figure}
\begin{center}
\includegraphics[width=4.0 in]{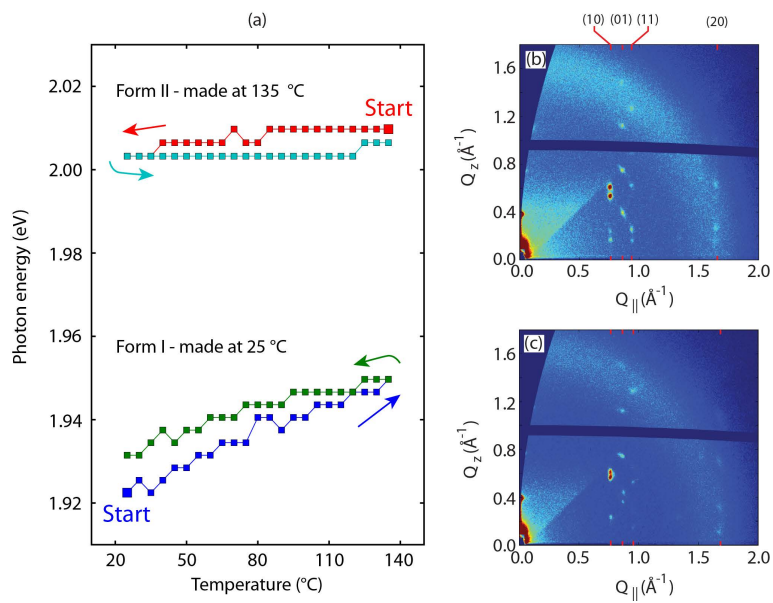}
\caption{\label{strained_peak_position} (a) In-situ reflection spectroscopy data for constrained  TIPS-pentacene thin films during cooling and heating cycles.  The upper data set is for a Form II film with  a thickness of about 26 nm, and the lower data set is for a 32 nm  Form I film.  These two films are listed as B and A respectively in Table \ref{tab:dep_params}.    Microscopy images of the Form I  and Form II sample after heating and cooling  are given in Supplementary Figs. S5 and  S6 respectively. (b,c) In-situ $\mu$GIWAXS of a Form II TIPS-pentacene thin film made at 135$^\circ$C,  on a Si/SiO$_2$ substrate.  This sample is referred to as D in Tables \ref{tab:dep_params} and \ref{tab:lattice_constants}.    (b) X-ray scattering pattern at the deposition temperature of  135$^\circ$C. (c) X-ray scattering pattern of the same film after cooling to 25$^\circ$C.}
\end{center}
\end{figure}

\begin{table}[ht]
\begin{center}
\caption{\label{tab:dep_params}Deposition conditions for isotropic thin films.  }
\begin{tabular*}{0.60\textwidth}{@{\extracolsep{\fill}}llccccc}
\hline \hline
Sample&Form&\mbox{T}&\mbox{Speed}&\mbox{Conc.}& Solvent & Thickness\\
~&~&\mbox{({$^\circ$C})}&\mbox{(mm/s)}&\mbox{(mg/ml)}& ~ &(nm)\\
\hline
A$^a$ & I  &  25 &     4 & 8    & Toluene   & 32\\
B$^a$ & II & 135 &    8 & 20  & Mesitylene & 26 \\
C$^b$  & I &   25 & 0.8 & 8.7 & Toluene   & 25 \\
D$^b$ & II & 135 &    4 & 22  & Mesitylene & 30 \\
\hline \hline 
\end{tabular*}
\begin{tabular*}{0.60\textwidth}{@{\extracolsep{\fill}}llccccc}
$^a$ Films A and B were deposited on glass substrates.  \\
$^b$ Films C and D were deposited on Si/SiO$_2$ substrates.
\end{tabular*}
\end{center}
\end{table}

In this section we report  thermal stability results for Form I and Form II films  during thermal cycling between 25$^\circ$C and 135$^\circ$C. The goal of this study is to observe whether the optical peaks can shift when lateral relaxation of the film is constrained by the substrate and  no cracking or buckling occurs. This situation can be modeled by adding a strain term to the free-energy equations that govern the phase transition, which favor the unstrained phase under a give set of conditions.\cite{kaganer2000strain}  In the present case, since we predetermine the strain by controlling the deposition temperature, we are able to engineer one or the other phase to be the unstrained phase.

The films are deposited under the conditions shown in Table \ref{tab:dep_params}. Very thin $\approx$30 nm films are obtained that have an isotropic in-plane structure composed of spherulitic grains. This type of grain structure is effectively a two-dimensional powder structure, which greatly simplifies grazing incidence wide-angle X-ray scattering (GIWAXS) measurements since  reflections from grains with many different in-plane orientations can be obtained in  a single exposure.   The films also have the advantage of being highly continuous  (see Fig. S7).  

In-situ optical reflection results are shown in Fig. \ref{strained_peak_position} for samples A and B. No cracks appear in either film after thermal cycling (see Supplementary Figs. S5 and  S6).  For the Form II sample, there is only a very small shift in the peak position during the cooling and heating cycle, indicating that Form II can be stabilized to 25$^\circ$C if no cracks occur to relieve the strain. The Form I films are  stable at room temperature indefinitely, and during annealing up to 135$^\circ$C.  A small reversible peak shift is observed when the sample is heated from 25$^\circ$C to 135$^\circ$C. Temperature cycles with longer annealing times at 135$^\circ$C  (not shown) have not revealed any additional peak shift -- that is, the peak position does not progressively shift with time.  It is conceivable that a small amount of reversible buckling occurs during heating.  Alternately, there could be a reversible rotation of the molecules within the unit cell, even though the overall lattice constants are constrained by the substrate.

The stability of Form I up to and beyond the bulk transition temperature at 124$^\circ$C is a key results since it is inconsistent with certain models for polymorph stabilization that postulate interface energetics or confinement effects as a mechanism for the stabilization.\cite{Diao2014JACS}     These models predict that the transition temperature  between Form II and Form I should be suppressed  further and further as the film thickness is reduced and interface energetics become dominant, and that Form II can be stabilized at room temperature for a thin enough film.  As a consequence,  Form I thin films should spontaneously transform to  Form II  well below the bulk transformation temperature, potentially even (eventually) at room temperature for very thin films.  In contrast, as we have already mentioned above, we do not observe such an effect in Fig. \ref{strained_peak_position}.   In addition, we have shown that there is no large nucleation barrier for the Form I to Form II transition, since it would cause a large hysteresis in the heating part of the temperature cycle, which we do not observe in Fig. \ref{thick_film_cracking}.  Thus, our results clearly confirm that strain is the only plausible mechanism behind polymorph stabilization of  TIPS-pentacene  thin films, in agreement with previous work.\cite{Ishviene2013JAP}

\begin{table}[ht]
\begin{center}
\caption{\label{tab:lattice_constants}In-plane lattice constants deduced from the grazing-incidence X-ray diffraction data.  }
\begin{tabular*}{0.56\textwidth}{llccccc}
\hline \hline
Sample&Form&T&\mbox{$a$}&\mbox{$b$}&\mbox{$\gamma$}&\mbox{$b \sin\gamma$}\\
~&~ &($^\circ$C)      &\mbox{({\AA})}&\mbox{({\AA})}&\mbox{(deg.)}&({\AA})\\
\hline
C&I  &25              & 7.72 & 7.76 & 80.7 & 7.65 \\
D&II &135$^a$    & 8.83 & 7.72 & 70.4 & 7.27 \\
D&II &25$^{b,c}$ & 8.72 & 7.71 & 71.0 & 7.29 \\
  \hline \hline \\
  \% Difference$^d$ &~&~& $+13.0$ & $-0.6$ & ~ & $-4.7$ \\
  \hline \hline
\end{tabular*}
\begin{tabular*}{0.60\textwidth}{llccccc}
$^a$ Made and measured at 135$^\circ$C.  \\
$^b$ Metastable 30 nm film made at 135$^\circ$C and measured at 25$^\circ$C.\\
 $^c$ Referred to as Form IIb by Diao $\textit{et al}$.$^6$  \\
$^d$ Comparing Form II (25$^\circ$C) to Form I (25$^\circ$C). The uncertainty is 0.5$\%$.
\end{tabular*}
\end{center}
\end{table}

In order to verify that the optical spectroscopy results correlate to structure, we have also performed X-ray diffraction experiments on TIPS-pentacene thin films.  Fig. \ref{strained_peak_position}(b) and (c) show in-situ grazing incidence X-ray scattering results for a Form II thin film (sample D in Tables \ref{tab:dep_params} and \ref{tab:lattice_constants}).  The film was deposited at  135$^\circ$C and then cooled to  25$^\circ$C.   Microscopy images in Supplementary Fig. S7 show that the film is composed of comet-shaped spherulitic grains, and that no cracks appeared as the film was cooled to 25 $^\circ$C.  Comparing Figs. \ref{strained_peak_position}(b) and (c), we see that the (10L), (01L), (11L), and (20L) reflections do not shift significantly in the in-plane (Q$_{||}$) direction.  This indicates that the crystalline lattice is constrained to the substrate.  However, some of the reflections shift along the Q$_z$ direction, which indicates that there is a vertical contraction or reorganization within the crystalline unit cell as the sample is cooled. These results are completely consistent with the optical spectroscopy results shown in Fig. \ref{strained_peak_position} for sample B, which suggest that there is almost no change in the molecular packing within each crystalline layer  as the Form II sample is cooled to 25$^\circ$C.

A Form I sample was fabricated at 25$^\circ$C (Sample C in Tables \ref{tab:dep_params} and \ref{tab:lattice_constants}) for comparison.   Grazing incidence X-ray scattering results for this sample are shown in Supplementary Fig. S8.  The in-plane lattice constants listed in Table \ref{tab:lattice_constants} are calculated from the Q$_{||}$ positions of the (10L), (01L) and (11L) reflections, which are themselves given in the Supplementary Table S2.  In Table \ref{tab:lattice_constants}, we see that Form II is expanded along the $a$-axis and it is contracted perpendicular to the $a$-axis relative to Form I.  This latter dimension is most accurately given by {$b \sin\gamma$. Most interestingly, we see that the in-plane lattice constants for the Form II sample (sample D) hardly change on cooling to 25$^\circ$C, indicating that the film is constrained to the substrate.  This confirms the conclusion of our optical study that the crystalline lattice is constrained to the substrate under conditions where no  cracking occurs. We also note that it has previously been found that the $a$-axis of aligned TIPS-pentacene thin films prepared by the pen writer method is  parallel to the deposition direction.\cite{wo2012JAP} The large positive thermal expansion along the $a$-axis in an aligned film means that cracks tend to occur perpendicular to the writing direction when samples are cooled from high temperature to room temperature.  As we have already noted, this observation explains the predominant direction of the cracks in aligned thin films, such as those shown  in Fig. \ref{cracking_images}.

\section*{Methods}

TIPS-pentacene ($\geq$ 99\%, Sigma Aldrich) solutions were prepared with either toluene (semiconductor grade, 99$\%$, Alfa Aesar) for room temperature deposition, or with mesitylene (98\%, Sigma-Aldrich) for all elevated temperature ($>25^\circ$C) depositions. Either silicon wafers [Figs. \ref{Expt_Layout}(c),  \ref{optical}(d,e)] or glass slides [Figs. \ref{equilibrium_peak_position}, \ref{thick_film_cracking}, and \ref{strained_peak_position}(a)] were used as substrates for thin films fabricated for spectroscopy measurements and for unpolarized microscopy. Silicon with a 300 nm thermally grown silicon oxide layer was also used as substrates for polarized microscopy experiments since the grain structure contrast is enhanced due to an optical interference effect produced by the oxide layer. The substrates were sonicated in toluene, acetone, and isopropanol for 7 min. and then treated with phenyltrichlorosilane (PTS, $\geq$ 97\% , Sigma Aldrich) to ensure proper wetting for the TIPS-pentacene solution. PTS treatment was accomplished by immersing the cleaned wafer or glass slide into a toluene solution of 3 wt\% PTS  and heated to 110$^\circ$C for 15 h under a Nitrogen overpressure to prevent evaporation of the solvent.  After PTS treatment, the substrates were sonicated in toluene, acetone and isopropanol for 1 min. each.

A polarizing optical microscope (Olympus BXFM) with ultra-long working distance objective lenses is used to observe the thin film growth in real-time. The system includes  an integrated UV-Vis spectrometer (Angstrom Sun Technologies Inc.) to acquire reflectance spectra over a selected small area, as shown in Fig. \ref{Expt_Layout}. Spectra were converted to absolute reflection percentage by comparion with a silicon mirror with a well-known reflectivity.  A silicon mirror was placed behind the glass substrates [Figs. \ref{equilibrium_peak_position}, \ref{thick_film_cracking}, and \ref{strained_peak_position}(a)], which serves to improve the signal and absorption contrast, as well as reducing peak shifts due to index of refraction dispersion.  In the hollow capillary writing process, deposition is carried out by allowing the solution held in a rectangular capillary to make contact with the substrate, followed by lateral translation of the substrate at a constant writing speed. \cite{wo2008APL} The sample is mounted on a thermoelectric module for temperature control. Heating at cooling rates are about 30$^\circ$C/min unless otherwise noted.
      
In-situ micro beam grazing incidence wide angle X-ray diffraction ($\mu$GIWAXS) was performed at the Cornell High Energy Synchrotron Source (CHESS) on the D1 beamline. X-rays with an energy of 10.74 keV  ($\lambda$ = 1.15 {\AA}) were focused to a beam size of 20 $\mu$m $\times$ 20 $\mu$m using a single-bounce X-ray capillary.{\cite{Detlef2013PSS} The incidence angle was 0.14$^\circ$ and the scattering patterns were recorded using a Pilatus 200K area detector with a pixel size of 172 $\mu$m. Si/SiO$_2$ wafers were used as substrates for all of the X-ray experiments. Film morphology and film thickness were characterized by atomic force microscopy (AFM).

%

\bibliography{reflectometry_peakshift_V12}
 
\section*{Acknowledgements}

The authors thank Arthur Woll for assistance with the X-ray diffraction experimental setup. This work was supported by the National Science Foundation, Division of Materials Research,  Electronic and Photonic Materials Program through award DMR-1307017. The X-ray scattering research was conducted at the Cornell High Energy Synchrotron Source (CHESS) which is supported by the NSF and the National Institutes of Health/National Institute of General Medical Sciences under NSF award DMR-1332208.

\section*{Author contributions statement}

Y.L. and R.H. conceived the experiments and wrote the manuscript, Y.L. J.W. and R.H.  conducted the experiments, and D.S. N.B. R.S. provided support and advice for specific measurements (X-ray diffraction, Atomic Force Microscopy, and Optical Spectroscopy, respectively).  All authors reviewed the manuscript. 

\section*{Additional information}

The authors declare no competing financial interests.

\end{document}